\documentclass[aps,pra,twocolumn,amsmath,amssymb,nofootinbib,showpacs,superscriptaddress]{revtex4-1}
\usepackage[english]{babel}
\usepackage{latexsym}
\usepackage{graphics}
\usepackage{graphicx}
\usepackage{epsfig}
\usepackage{color}
\usepackage{bm}
\usepackage{amsmath}
\usepackage{amssymb}
\usepackage{amsthm}
\usepackage{dcolumn}
\usepackage{bm}
\usepackage{float}
\usepackage{hyperref}
\usepackage{color}
\usepackage{epstopdf}
\usepackage{braket}
\usepackage{physics}
\usepackage{cleveref}
\usepackage{courier}
\usepackage{mathtools}
\usepackage{bbold}
\usepackage[normalem]{ulem}
\usepackage[svgnames]{xcolor}
\usepackage{tikz}
\usetikzlibrary{quantikz}

\hypersetup{hidelinks,colorlinks=true,allcolors=DarkBlue}

\def\myvdots{\ \vdots\ }
\theoremstyle{remark}

\begin{document}
\preprint{APS/123-QED}
\title{Suppressing decoherence in quantum state transfer with unitary operations}
	
	\author{M.A. Gavreev}
	\affiliation{Russian Quantum Center, Skolkovo, Moscow 143025, Russia}
	\affiliation{National University of Science and Technology ``MISIS”, Moscow 119049, Russia}
	
	\author{E.O. Kiktenko}
	\affiliation{Russian Quantum Center, Skolkovo, Moscow 143025, Russia}
	\affiliation{National University of Science and Technology ``MISIS”, Moscow 119049, Russia}
	
	\author{A.S. Mastiukova}
	\affiliation{Russian Quantum Center, Skolkovo, Moscow 143025, Russia}
	\affiliation{National University of Science and Technology ``MISIS”, Moscow 119049, Russia}
	
	\author{A.K. Fedorov}
	\affiliation{Russian Quantum Center, Skolkovo, Moscow 143025, Russia}
	\affiliation{National University of Science and Technology ``MISIS”, Moscow 119049, Russia}

\date{\today}
\begin{abstract}
Decoherence is the fundamental obstacle limiting the performance of quantum information processing devices. 
The problem of transmitting a quantum state (known or unknown) from one place to another is of great interest in this context. 
In this work, by following the recent theoretical proposal [Opt. Eng. {\bf 59}, 061625 (2020)] 
we study an application of quantum state-dependent pre- and post-processing unitary operations for protecting the given (multi-qubit) quantum state against the effect of decoherence acting on all qubits.
We observe the increase in the fidelity of the output quantum state both in a quantum emulation experiment, 
where all protecting unitaries are perfect, and in a real experiment with a cloud-accessible quantum processor, where protecting unitaries themselves are affected by the noise.
We expect the considered approach can be useful for analyzing capabilities of quantum information processing devices in transmitting known quantum states.  
We also demonstrate an applicability of the developed approach for suppressing decoherence in the process of distributing a two-qubit state over remote physical qubits of a quantum processor.
\end{abstract}

\maketitle

\section{Introduction}

During recent decades, remarkable progress in developing quantum information processing devices has been performed~\cite{Wehner2018}. 
Specifically, currently available quantum processing devices are able to solve computational problems close to limits of what can be done with most powerful classical technologies~\cite{Fedorov2022,Martinis2019,Pan2020}. 
However, available quantum information processing devices have serious limitations, whose fundamental reason is decoherence. 
As a result, the number of operations that can be implemented before the level of errors exceeds the critical level is modest~\cite{Babbush2021-4}.  
The possibility to achieve substantial computational advances in practically relevant problems with noisy intermediate-scale quantum (NISQ) devices is questionable~\cite{Fedorov2022}, 
so the path to useful quantum advantage requires the noise suppression. 
A possible solution that addresses this problem is to use quantum error correction~\cite{Shor1995,Shor1996,Shor1996-2,Bacon2006,Martinis2012,Antipov2022}.
In the last decade, principles of error correction schemes in the quantum domain have been demonstrated in experiments with optical systems~\cite{Franson2005}, 
trapped ions~\cite{Wineland2004,Blatt2011,Blatt2020,Monroe2021,Blatt2021-2}, 
neutral atoms~\cite{Lukin2022},
and superconducting qubits~\cite{Schoelkopf2012,Schoelkopf2016,Kelly2021,Pan2021-3,Acharya2022}. 
In addition to quantum error correction, one may consider error suppression that would reduce, but not fully eliminate, the effect of decoherence during processing a quantum state~\cite{Dieter2016}.
A variety of methods have been considered to achieve this goal~\cite{Gambetta2017,Gambetta2019,Benjamin2017,Benjamin2018,Lidar2014,Lidar2017-2,Lidar2018-4}. 
Albeit applying error suppression techniques is of help to extend computational capabilities of quantum devices, 
the experimental challenges for their practical implementation require additional progress in studies of the workflow of quantum processors~\cite{Gambetta2019}.

A specific problem that appears within the context of quantum information processing is how to transmit a quantum state (known or unknown) from one place to another taking into account the effect of decoherence. 
The importance of this problem, especially in the case of known quantum states, lies in the fundamental domain of quantum information science, 
although various schemes for its usage in quantum data buses between quantum registers and/or processors capable of transmitting arbitrary quantum states have been discussed~\cite{Bose2007}.
Considerable attention has been paid to the theoretical analysis of using quantum spin chains for the purpose of quantum state transfer~\cite{Bose2003,Ekert2004,Bose2007,Roncaglia2007,Smith2012,Sousa2014,Feldman2021}.
Such an interest can be explained by the possibility to transfer (known or unknown) quantum states without additional interfaces~\cite{Bose2003}.
The question related to the role of decoherence in this case still remains crucial in this scope of tasks and generally for the field of quantum information~\cite{Saffman2010,Lloyd2012,Diep2020}.

\begin{figure}[t]
	\begin{centering}
		\includegraphics[width=0.75\columnwidth]{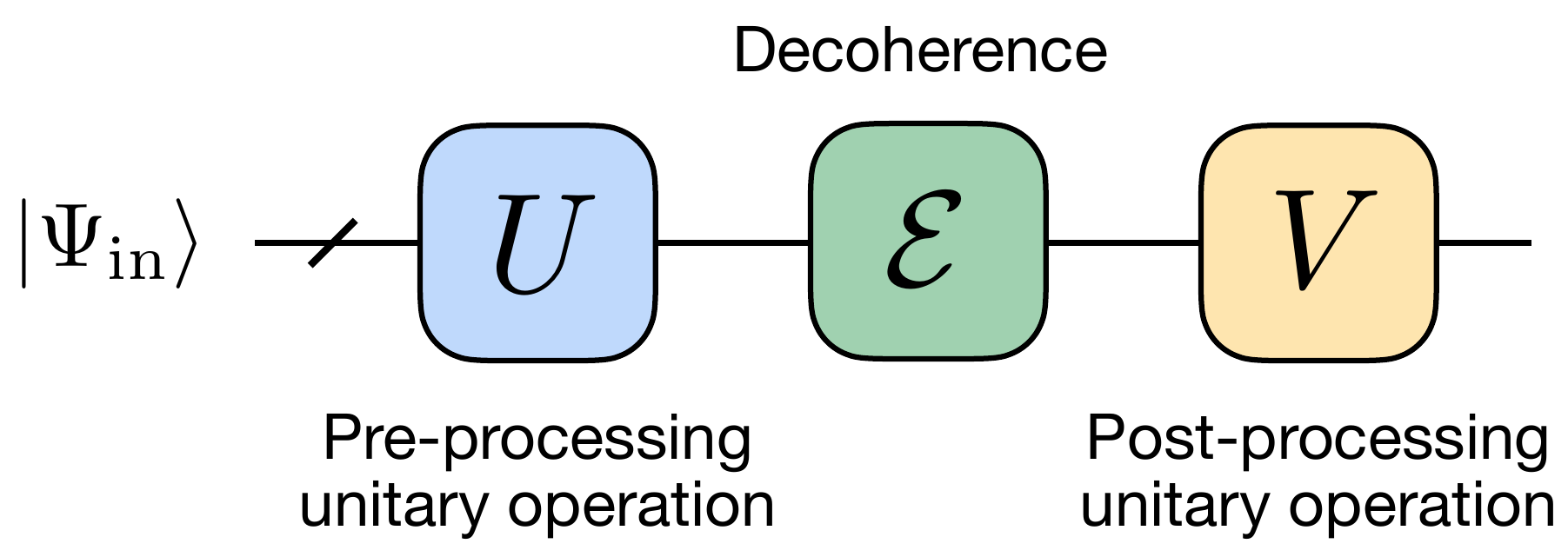}
	\end{centering}
	\vskip -3mm
	\caption{Quantum circuit of error suppression based on pre-processing and post-processing unitary operations.}
	\label{fig:genscheme}
\end{figure}

In this work, we stress a particular scheme for suppressing the effect of decoherence, which is based on the use of unitary operation.
This approach has been recently proposed theoretically in Ref.~\cite{Kiktenko2020}. 
The general idea is to surround the decoherence channel ${\cal E}$ by two unitary operators $U$ and $V$ (see Fig.~\ref{fig:genscheme}), 
which we refer to as pre- and post-processing, correspondingly, whose form is determined both by the decoherence channel ${\cal E}$ and protected state $\ket{\Psi_{\rm in}}$.
In contrast to the previous work~\cite{Kiktenko2020}, where ${\cal E}$ was assumed to be acting non-trivially on one particular qubit, here we consider a more general case, where all qubits are affected by the decoherence process.
First, we demonstrate a possibility to suppress decoherence effects in a quantum emulation experiment, 
where all qubits are affected by the same depolarizing, dephasing, or amplitude damping channels, and an implementation of all unitary gates is assumed to be ideal.
We then implement the same protection scheme on a cloud-accessible quantum processor, where the used protecting gates are imperfect by themselves.
In this case, we find that the protection scheme starts providing an advantage in the output fidelity starting from a certain threshold of the decoherence strength.
Finally, we show how our scheme can be used for protecting two-qubit states during its distribution over remote physical qubits of a quantum processor, i.e. qubits which can not be connected directly by native two-qubit gates.

Our paper is organized as follows.
In Sec.~\ref{sec:scheme}, we describe a general scheme of applying unitary operations for suppressing decoherence. 
In Sec.~\ref{sec:emulators}, we apply the scheme for emulators of quantum computers and analyze the efficiency of various schemes. 
In Sec.~\ref{sec:processors}, we implement the same protection scheme on a cloud-accessible quantum processor.
In Sec.~\ref{sec:state-transfer}, we analyze the applicability of the scheme for the case of quantum state transfer.
We conclude in Sec.~\ref{sec:outlook}

\section{Error suppression using unitary operations}\label{sec:scheme}

Let us consider a system of $n$ two-level particles (qubits) initially prepared in a joint (perhaps, entangled), pure state $\rho_{\rm in}=\ket{\Psi_{\rm in}}\bra{\Psi_{\rm in}}$.
Due to an uncontrolled interaction with an external environment, this state suffers from a decoherence, which we describe by a completely positive trace-preserving (CPTP) map $\mathcal{E}[\cdot]$.
In what follows, we refer to ${\cal E}$ as a \emph{decoherence channel}. 
The destructive effect of the decoherence channel can be quantified by fidelity
\begin{equation} \label{eq:basic-fidelity}
    F_{\rm W/O} := \bra{\Psi_{\rm in}} \mathcal{E} [\rho_{\rm in}]\ket{\Psi_{\rm in}} \leq 1.
\end{equation}
Here subscript ${\rm W/O}$ indicates that this is the default case without applying the scheme described below.

In our work, we study possibilities to suppress decoherence, and thus increase the fidelity, by applying additional transformations to the system.
We take these transformations in the form of additional unitary operators, $U$ and $V$, named \emph{pre-processing} and \emph{post-processing} gates, and acting just before and after the action of the decoherence channel correspondingly.
Importantly, $U$ and $V$ are designed specifically for given initial state $\rho_{\rm in}$ and decoherence channel ${\cal E}$.
The expression for the resulting fidelity is given by the following expression:
\begin{equation}
    F_{U,V}= \bra{\Psi_{\rm in}} V \mathcal{E} \left[ U \rho_{\rm{in}} U^{\dagger} \right] V^{\dagger} \ket{\Psi_{\rm in}}.
\end{equation}
In our consideration, we take $U$ and $V$ from certain, generally restricted, sets of unitaries ${\cal U}$ and ${\cal V}$, respectively.
One can see that as long as ${\cal U}$ and ${\cal V}$ contain identity operators,
\begin{equation}
    \max_{U\in{\cal U},V\in{\cal V}} F_{U,V} \geq F_{\rm W/O},
\end{equation}
and so we expect an increase in fidelity compared to the default case~\eqref{eq:basic-fidelity}.

\begin{figure}[t]
	\begin{centering}
	\includegraphics[width=1\columnwidth]{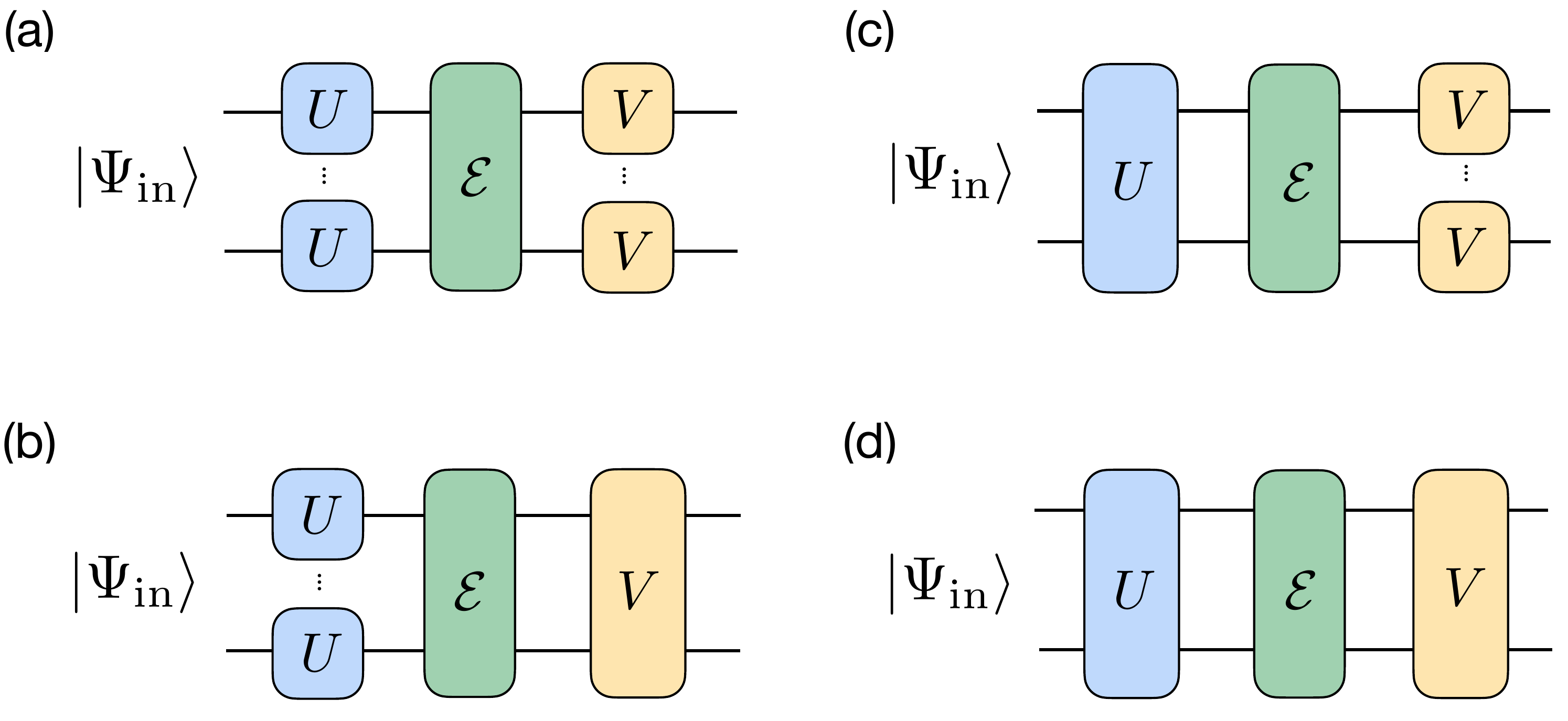}
	\end{centering}
	\vskip -5mm
	\caption{Schemes for protecting a pure state $\ket{\Psi_{\rm {in}}}$ from the decoherence channel $\mathcal{E}$:
	individual-individual (a), individual-collective (b), collective-individual scheme (c), and collective-collective (d).
	Note that single-qubit $U$ and $V$ can be different for different qubits.}
	\label{fig:scheme2}
\end{figure}

We consider two types of pre- and post-processing unitary operations: 
(i) \emph{individual} operations, consisting of single-qubit unitaries, i.e. taken from ${\rm U}(2)^{\otimes n}$ group, and (ii) \emph{collective} operations, represented as unitaries taken from the whole ${\rm U}(2^n)$ group.
Since each of the pre- and post-processing operators can be either individual or collective, we get four possible combinations: 
(a) both unitary operations are individual; 
(b) pre-processing unitary operation is individual, while the post-processing unitary operation is collective; 
(c) pre-processing unitary operation is collective, while the post-processing unitary operation is individual; 
and (d) both unitary operations are collective.
All four schemes are schematically shown in Fig.~\ref{fig:scheme2}.
Since (b) and (c) include (a) as a special case, and also (d) includes (a-c), the maximal achievable fidelities in the schemes satisfy the following inequality:
\begin{equation}
    \begin{aligned}
        &F_{\rm ind,ind} \leq \min(F_{\rm col,ind}, F_{\rm ind,col}) \\
        &\max(F_{\rm col,ind}, F_{\rm ind,col}) \leq F_{\rm col,col},
    \end{aligned}
\end{equation}
where the first (second) subindex specifies the type of the pre- (post-) processing operator.
We note that although replacing individual operators with collective ones leads to an increase in performance, constructing collective operators yields an additional overhead in the number of gates and the corresponding circuit depth: Individual operators have a unit-depth and consist of no more than $n$ single-qubit gates, while a transformation between two given pure $n$-qubit states, which is an aim of collective operators, generally requires an exponential in $n$ number of single- and two-qubit gates or an exponential number of ancillary qubits for linear depth circuits~\cite{plesch2011quantum, zhang2022quantum}.

The subject of how $F_{\rm ind,col}$ and $F_{\rm col,ind}$ are related to each other is rather intriguing. 
In Ref.~\cite{Kiktenko2020}, it has been demonstrated that for a single-qubit decoherence, i.e. 
\begin{equation}
	{\cal E}={\cal E}^{(1)}\otimes {\rm Id}^{\otimes (n-1)},
\end{equation} 
where ${\cal E}^{(1)}$ and ${\rm Id}$ are arbitrary and identity single-qubit channels correspondingly, one has $F_{\rm col,ind}=F_{\rm ind,col}$.
The maximal achievable fidelities for individual-collective and collective-individual schemes are also the same for an arbitrary self-dual ${\cal E}$.
Indeed, since
\begin{multline}
    F_{U,V}={\rm Tr}[{\cal E}[U\rho_{\rm in}U^\dagger] V^\dagger \rho_{\rm in}V]=\\={\rm Tr}[{\cal E}[V^\dagger \rho_{\rm in}V]U\rho_{\rm in}U^\dagger ]=F_{V^\dagger, U^\dagger},
\end{multline}
if a certain value of fidelity is achieved in one scheme, the same value can be achieved in the other scheme by replacing $U\rightarrow V^\dagger$, $V\rightarrow U^\dagger$.
Therefore the maximal achievable values are the same.
We leave the consideration of the relation between $F_{\rm col,ind}$ and $F_{\rm ind,col}$ in the case of a general decoherence channel for further studies.

\begin{figure}
\begin{equation*}
\mathcal{\mathcal{U}_{\rm prep}}(\theta) = 
\begin{quantikz}[row sep={0.6cm,between origins}, column sep=0.2cm]
& \gate{R_y(\theta)} &\ctrl{1} &\qw      &\push{~\ldots~}  & \qw     &\gate{H} &\qw \\ 
&        \qw         &\targ{}  &\ctrl{1} &\push{~\ldots~}  & \qw     &\gate{H} &\qw \\ 
&        \qw         &\qw      &\targ{}  &\push{~\ldots~}  & \qw     &\gate{H} &\qw \\ 
&  &  &  &  \lstick{\myvdots} &  &  &\\
&        \qw         &\qw      &\qw      &\push{~\ldots~}  &\ctrl{1} & \gate{H}&\qw \\ 
&        \qw         &\qw      &\qw      &\push{~\ldots~}  &\targ{}  & \gate{H}&\qw \\ 
\end{quantikz}
\end{equation*}
\vskip -7mm
\caption{Quantum circuit for the unitary operator performing preparation of the $n$-qubit input state $\ket{\Psi_{\rm in}(\theta)}$.}
\label{fig:prep-circ}
\end{figure}
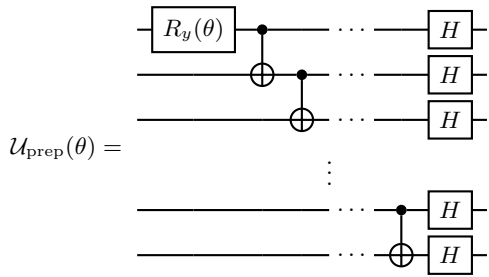

In what follows, we analyze the performance of different schemes with respect to the fixed state
\begin{equation} \label{eq:instate}
    \ket{\Psi_{\rm {in}}(\theta)} = \cos{\frac{\theta}{2}}\ket{+}^{\otimes n}+\sin{\frac{\theta}{2}}\ket{-}^{\otimes n}
\end{equation}
where $\ket{\pm}:=\frac{1}{\sqrt{2}}(\ket{0}\pm\ket{1})$ and we set $\theta=2\pi/3$.

Our choice for this state is motivated by several reasons. 
The first is that it is entangled, but it is not a maximally-entangled state.
According to the results of Ref.~\cite{Kiktenko2020}, it is the most interesting case for unitaries-based protection.
The second reason is that the reduced single-qubit states of $\ket{\Psi_{\rm {in}}(\theta)}$ are not diagonal in the computational basis.
It makes it interesting to consider these schemes concerning `basis-dependent' decoherence channels, e.g., amplitude damping and dephasing.
Third, $\ket{\Psi_{\rm {in}}(\theta)}$ can be easily prepared from $\ket{0}^{\otimes n}$ by applying ${\cal U}_{\rm prep}(\theta)$ gate, whose decomposition into standard single- and two-qubit gates is shown in Fig.~\ref{fig:prep-circ}.
Here and after, standard notations for controlled-NOT (CNOT), Hadamard, and Pauli rotation gates are used.
Next, we use ${\cal U}_{\rm prep}(\theta)$ as a template for constructing collective protecting operators.

Before proceeding, we would like to note that although from the practical point of view, the most interesting are individual-individual and collective-individual schemes, we consider all four scenarios for revealing the whole picture.
We also note that in the presented consideration the pre- and post-processing operations are assumed to be perfect.
However, this may not be the case at all in real-world setups~\cite{Gambetta2017,Gambetta2019,Benjamin2017,Benjamin2018,Lidar2014,Lidar2017-2,Lidar2018-4}.

\section{Demonstrating error suppression with a quantum emulator}\label{sec:emulators}

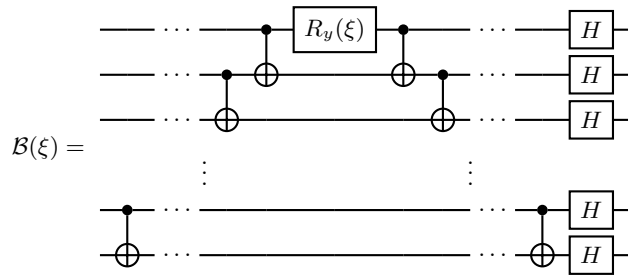
\begin{figure}[t]
\def\myvdots{\ \vdots\ }
\begin{equation*}
\mathcal{\mathcal{B}}(\xi) = 
\begin{quantikz}[row sep={0.6cm,between origins}, column sep=0.2cm]
&\qw      &\push{~\ldots~}&\qw              &\ctrl{1}&\gate{R_y(\xi)} &\ctrl{1} &\qw      &\push{~\ldots~}     &\qw      &\gate{H} &\qw \\ 
&\qw      &\push{~\ldots~}&\ctrl{1}         &\targ{} &\qw             &\targ{}  &\ctrl{1} &\push{~\ldots~}     &\qw      &\gate{H} &\qw \\ 
&\qw      &\push{~\ldots~}&\targ{}          &\qw     &\qw             &\qw      &\targ{}  &\push{~\ldots~}     &\qw      &\gate{H} &\qw \\ 
&         &            &\lstick{\myvdots}&        &                &         &         &\lstick{\myvdots}&         &         & \\
&\ctrl{1} &\push{~\ldots~}&\qw              &\qw     &\qw             &\qw      &\qw      &\push{~\ldots~}     &\ctrl{1} &\gate{H} &\qw \\ 
&\targ{}  &\push{~\ldots~}&\qw              &\qw     &\qw             &\qw      &\qw      &\push{~\ldots~}     &\targ{}  &\gate{H} &\qw \\ 
\end{quantikz}
\end{equation*}
\vskip -7mm
\caption{Quantum circuit of the collective operation $\mathcal{B}(\xi)$.}
\label{fig:collective-circ}
\end{figure}

Here we demonstrate the performance of the considered error suppression scheme using an emulator of a quantum processor.
Namely, we employ \texttt{AerSimulator} emulator provided as part of the \texttt{qiskit} package~\cite{Aer}. 
We consider a simplified error model, where the decoherence channel is taken as a tensor power of a single-qubit channel ${\cal E}^{(1)}$, namely,
\begin{equation}
    {\cal E}[\cdot] = \left({\cal E}^{(1)}\right)^{\otimes n}[\cdot].
\end{equation}
We also assume the ideal realization of employed pre- and post-processing unitaries.

Three basic decoherence channels are studied: (i) amplitude damping, (ii) dephasing, and (iii) depolarizing one.
Their Kraus operators (${\cal E}^{(1)}[\cdot]=\sum_k A_k \cdot A_k^\dagger$) are respectively defined as
\begin{eqnarray}
    &A_1 = \ket{0}\bra{0} + \sqrt{1-p}\ket{1}\bra{1}, \quad 
    A_2 = \sqrt{p}\ket{0}\bra{1}, \label{eq:kraus_ad}\\
    &A_1 = \frac{\sqrt{p}}{2}(\mathbb{1}+\sigma_3), \quad
    A_2 = \frac{\sqrt{p}}{2}(\mathbb{1}-\sigma_3), \label{eq:kraus_deph}\\
    &A_0 = \sqrt{1-\frac{3p}{4}}\mathbb{1}, \quad A_i=\sqrt{\frac{p}{4}}\sigma_i \quad (i=1,2,3),
\end{eqnarray}
where $\mathbb{1}$ is a single-qubit identity matrix, $\sigma_1$, $\sigma_2$, $\sigma_3$ stand for standard $x$-, $y$-, $z$ Pauli matrices, and an additional parameter $p\in[0,1]$ determines a decoherence strength.
We note that the form of the amplitude damping channel~\eqref{eq:kraus_ad} corresponds to a spontaneous decay from $\ket{1}$ to $\ket{0}$, dephasing channel corresponds to destruction of non-diagonal elements, 
and depolarizing is `basis-invariant' in a sense that ${\cal E}[u\cdot u^\dagger]=u{\cal E}[\cdot]u^\dagger$ for any unitary $u$.
We also note that the dephasing and depolarizing channels are self-dual, since their Kraus operators are Hermitian.

In what follows, we discuss the choice of unitaries for each type of protection scheme. 
The idea is that $U$ and $V$ should meet two conditions: 
(i) after the action of $U$, the input state  should be maximally robust to noise, and (ii) the action of $U$ and $V$ together should preserve the input state as much as possible. 
We choose unitaries for each protection scheme based on these conditions.

In the case of individual-individual scheme, we take
\begin{equation}
    U = (XH)^{\otimes n}, \quad V = (HX)^{\otimes n},
\end{equation}
where $H$ is a standard Hadamard gate, and $X=\sigma_1$ is a $\pi$ rotation around $x$-axis.
The idea behind this choice is in turning each of reduced single-qubit states of $\ket{\Psi}_{\rm in}$ into a diagonal form s.t. the population on $\ket{0}$ is larger than the one on $\ket{1}$.
This form of a density matrix is the most robust against considered decoherence channels~\cite{Kiktenko2020}.

\begin{figure}[t]
	\begin{centering}
		\includegraphics[width=1.\columnwidth]{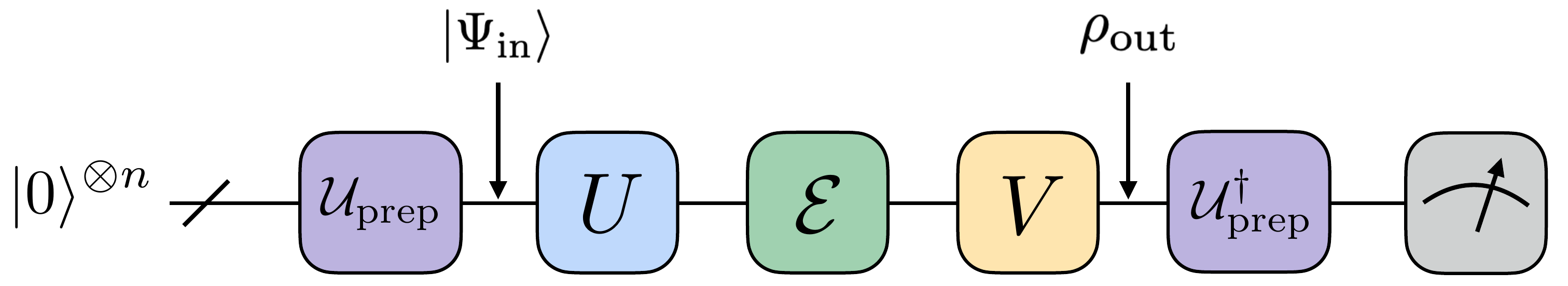}
	\end{centering}
	\vskip -5mm
	\caption{Circuit for the measurement of the resulting fidelity. 
	Parameter $\theta$ for the ${\cal U}_{\rm prep}$ is taken to be equal to $2\pi/3$.}
	\label{fig:meas}
\end{figure}

\begin{figure*}[t]
	\begin{centering}
	\includegraphics[width=1.8\columnwidth]{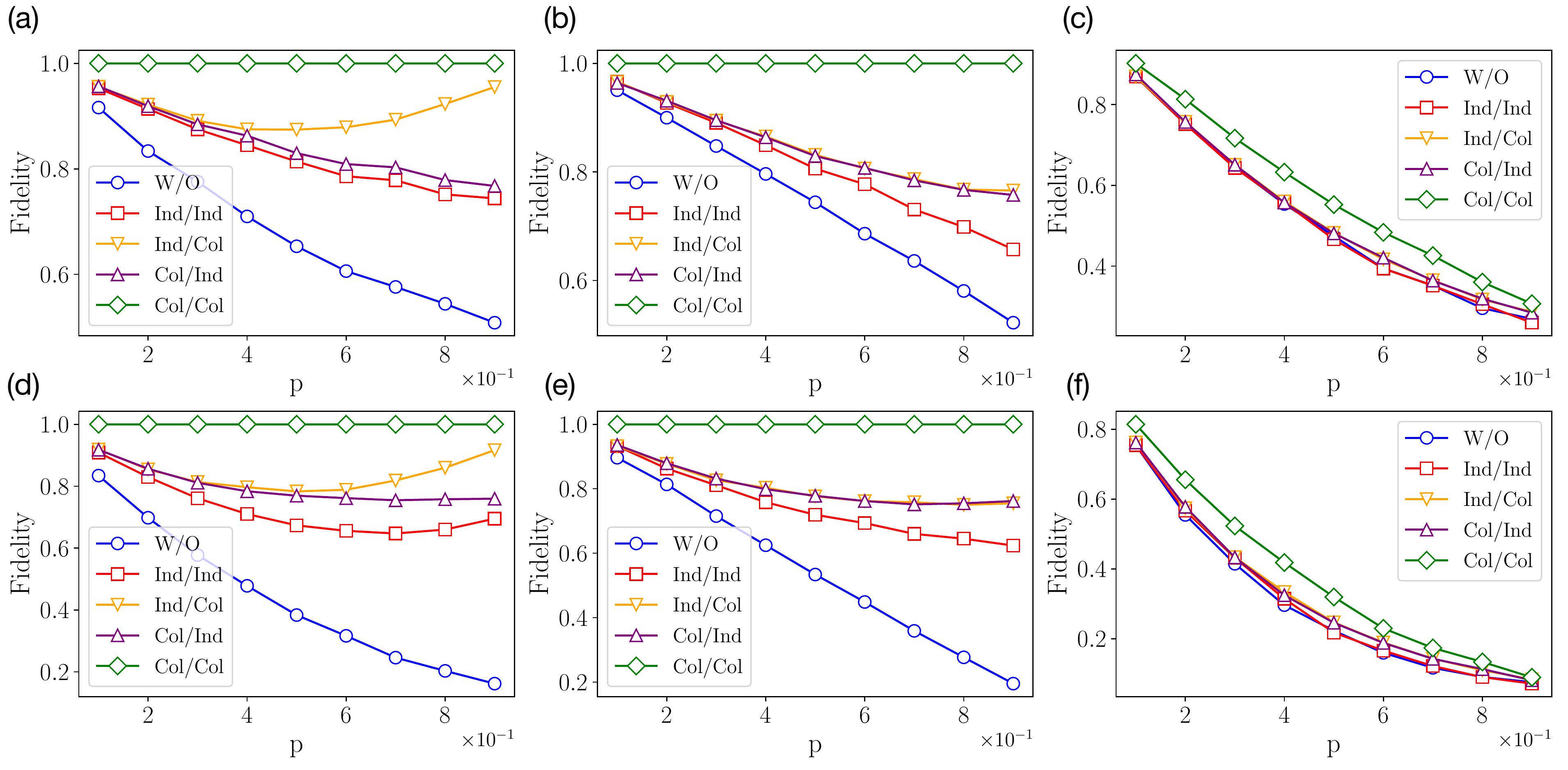}
	\end{centering}
	\caption{
	Simulation results for $n=2$ qubit (a-c) and $n=4$ qubit (d-f) cases.
	The amplitude damping (a,d), dephasing (b,e), and depolarizing (c,f) channels are considered.}
	\label{fig:sim-1}
\end{figure*}

In the case of individual-collective scheme, we take
\begin{equation}
    U = (XH)^{\otimes n}, \quad V = {\cal B}(\xi_{\rm opt})X^{\otimes n},
\end{equation}
where ${\cal B}(\xi)$ is a unitary whose circuit is shown in Fig.~\ref{fig:collective-circ}, and the angle $\xi_{\rm opt}$ is obtained by maximizing resulting fidelity value
\begin{equation}
 |\bra{\Psi_{\rm in}} \mathcal{B}(\xi) X^{\otimes n} \mathcal{E} \left[ U \ket{\Psi_{\rm in}} \bra{\Psi_{\rm in}} U^{\dagger} \right] X^{\otimes n} \mathcal{B}^{\dagger}(\xi)  \ket{\Psi_{\rm in}}|^2
\end{equation}
over $\xi\in[0, 2\pi]$.
The choice of the collective operator's form is motivated by the state preparation circuit.
In fact, it corresponds to a deconstruct-construct sequence with an additional rotation in the middle (defined by parameter $\xi$) corresponding to a change of an entanglement parameter initially specified by $\theta$ in the state preparation circuit.

The case of collective-individual scheme is similar to individual-collective one, and is obtained by taking
\begin{equation}
    U=X^{\otimes n}{\cal B}^\dagger(\xi_{\rm opt}'), \quad
    V=(HX)^{\otimes n},
\end{equation}
where $\xi_{\rm opt}'$ is obtained by maximizing
\begin{equation}
    |\bra{\Psi_{\rm in}} V \mathcal{E} \left[ X^{\otimes n} \mathcal{B}^{\dagger}(\xi) \ket{\Psi_{\rm in}} \bra{\Psi_{\rm in}} \mathcal{B}(\xi) X^{\otimes n} \right] V^\dagger \ket{\Psi_{\rm in}}|^2.
\end{equation}

Finally, the collective-collective scheme is obtained by taking 
\begin{equation}
    U = {\cal U}_{\rm prep}(\theta)^\dagger, \quad 
    V = {\cal U}_{\rm prep}(\theta)
\end{equation}
with $\theta=2\pi/3$.
This choice corresponds to deconstruction of $\ket{\Psi_{\rm in}}$ back to $\ket{0}^{\otimes n}$, which is the most stable state with respect to the considered decoherence channels, and re-preparation of $\ket{\Psi_{\rm in}}$ after the action of channel.

To measure fidelity values, we implement the circuit shown in Fig.~\ref{fig:meas} for two- and four-qubit quantum systems.
We consider four protection schemes, as well as a default scheme without protection, with respect to the three introduced types of decoherence channels, and different values of $p$.
Simulation results are obtained for $N=10000$ runs of each circuit with fixed decoherence channel (and fixed value of $p$).
Fidelity values are calculated as frequencies of obtaining all-zeros outcomes in read-out measurements.
The results are shown in Fig.~\ref{fig:sim-1}.
First of all, we note that the following inequalities hold for all decoherence models:
\begin{equation}
    F_{\rm W/O} \leq F_{\rm ind,ind} \leq F_{\rm col,ind} \leq F_{\rm ind, col} \leq F_{\rm col,col}.
\end{equation}
Moreover, $F_{\rm ind,col}=F_{\rm col,ind}$ for self-dual dephasing and depolarizing channels, and $F_{\rm ind,ind}=F_{\rm W/O}$ for covariant depolarizing channel.
We also note that $F_{\rm col,col}=1$ for amplitude damping and dephasing channels, since $\ket{0}^{\otimes n}$ is not affected by the decoherence, yet $F_{\rm col,col}<1$ for the depolarizing channel.
The reason for the latter fact is that depolarizing channel always increases mixedness of input states, and there is no pure state that is preserved under the depolarizing.
Therefore, even collective-collective scheme does not provide unit fidelity in the case of depolarizing channels.

There is also an interesting effect of non-monotonic behaviour of fidelity as a function of $p$ in the case of amplitude damping channel in individual-collective and individual-individual schemes.
In the case of the individual-collective scheme, this behaviour can be explained by the fact the post-processing collective operator actually serves for re-preparing the state $\ket{\Psi_{\rm in}}$ from the state $\rho$ coming from $\cal{E}[\cdot]$.
The resulting fidelity is thus determined by the purity of $\ket{\rho}$.
In the case of $p=1$, amplitude damping channel outputs pure $\rho=(\ket{0}\bra{0})^{\otimes n}$, and so we achieve the unit fidelity.

\begin{figure*}[ht!]
		\begin{centering}
			\includegraphics[width=1.4\columnwidth]{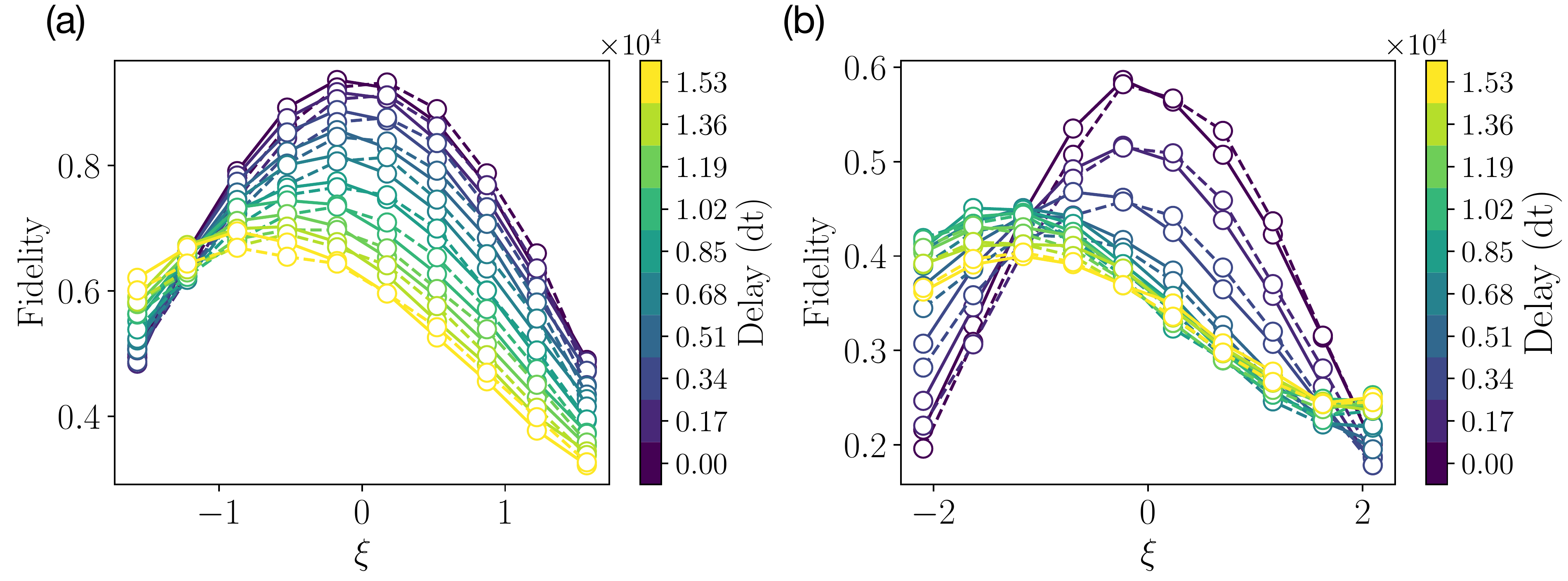}
		\end{centering}
		\caption{Experimental results of $\xi$ angle calibration from \textrm{IBMQ} quantum processor for (a) two-qubit experiment; (b) four-qubit experiment. 
		Solid and dashed lines stand for collective-individual and individual-collective schemes respectively.}
		\label{fig:ibmq-calibration}
\end{figure*}

\begin{figure*}[ht!]
		\begin{centering}
			\includegraphics[width=1.3\columnwidth]{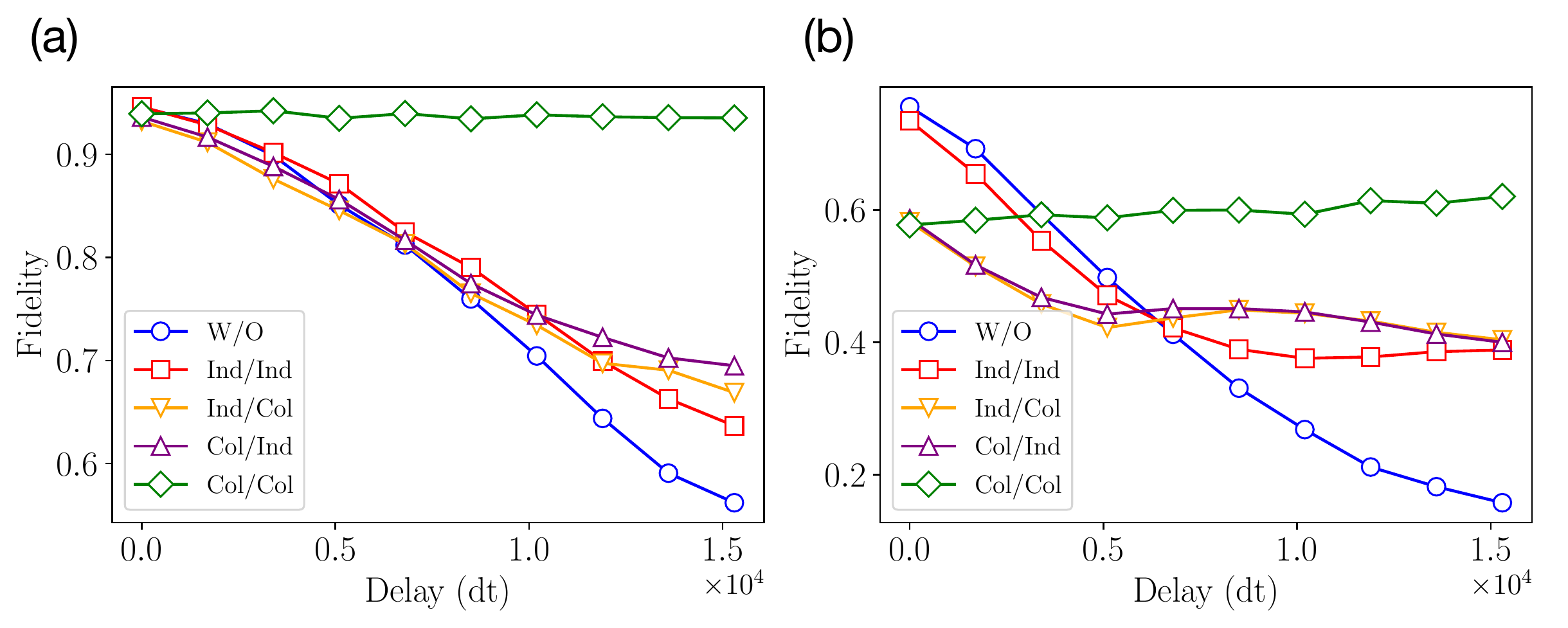}
		\end{centering}
		\caption{Experimental results from \textrm{IBMQ} quantum processor for (a) two-qubit experiment; (b) four-qubit experiment.}
		\label{fig:ibmq-results}
\end{figure*}

For the individual-individual scheme, the resulting fidelity takes the following form:
\begin{equation}
    F_{\rm ind,ind} = \bra{{\chi}} \mathcal{ E}\left[\ket{{\chi}}\bra{{\chi}}\right] \ket{{\chi}},
\end{equation}
where $\ket{\chi}=c\ket{0}^{\otimes n}+s\ket{1}^{\otimes n}$ with $c:=\cos{(\pi/6)}$, $s:=\sin{(\pi/6)}$.
For $\cal{E}$ in the form of single-qubit amplitude damping channels of the same strength $p$, it transforms into
\begin{equation}
    F_{\rm ind,ind} = c^4+p^ns^2c^2 + (1-p)^ns^4+2(1-p)^{p/2}c^2s^2,
\end{equation}
which is non-monotonic for $n>2$.

\section{Validating error suppression with a cloud-based quantum processor}\label{sec:processors}

In this section, we implement circuits from the previous section on a cloud-accessible 5-qubit quantum processor \texttt{ibmq\char`_manila}. 
To access the decoherence process on the real device, we utilize the \texttt{Delay} instructions in the natural time units (\texttt{dt}). 
Delay time can be seen as the strength of the decoherence distortion of the input state on the real device.

It is also important to note that the form of the decoherence channel is not exactly known.
One can expect that ${\cal E}[\cdot]$ has a tensor product form of single-qubit channels, each consisting of depolarizing, dephasing, amplitude damping ones~\cite{georgopoulos2021modeling}.
However, certain collective decoherence effects also can take place due to a qubits' crosstalk.
That is why to optimize the parameter of the collective protecting operation $\mathcal{B}(\xi)$, we perform additional measurements. 
We demonstrate the resulting dependence of the input-output fidelity on the angle $\xi$ in Fig.~\ref{fig:ibmq-calibration}. 
We observe the drift of the optimal parameter $\xi$ with the increase of delay time. 
This highlights the usefulness of utilizing the collective operator in this case.

The results of the final fidelity measurements for $n=2$ and $n=4$ qubits are shown in Fig. \ref{fig:ibmq-results}. 
We note that the resulting input-output fidelities, given in this section, are obtained experimentally without additional circuit transpilation. 
We can see that the resulting input-output fidelity is lower for any protection scheme compared with unprotected case for weak decoherence action.
This is due to the additional noise from single- and multiqubit gates in pre- post- processing operators.
However, with the enhancement of the decoherence, we observe the gain from protection schemes. 
This demonstrates the existence of a tradeoff between the use of additional unitary operations to prevent decoherence and the noise created by these operations.
Similar to the results of the previous section, the collective-collective scheme shows the best performance starting from some large enough noise. 
This can be explained by the triviality of the protection operators action. 
At the same time, we observe that individual-collective and collective-individual schemes are almost of the same performance.

\begin{figure}[ht!]
		\begin{centering}
			\includegraphics[width=1.\columnwidth]{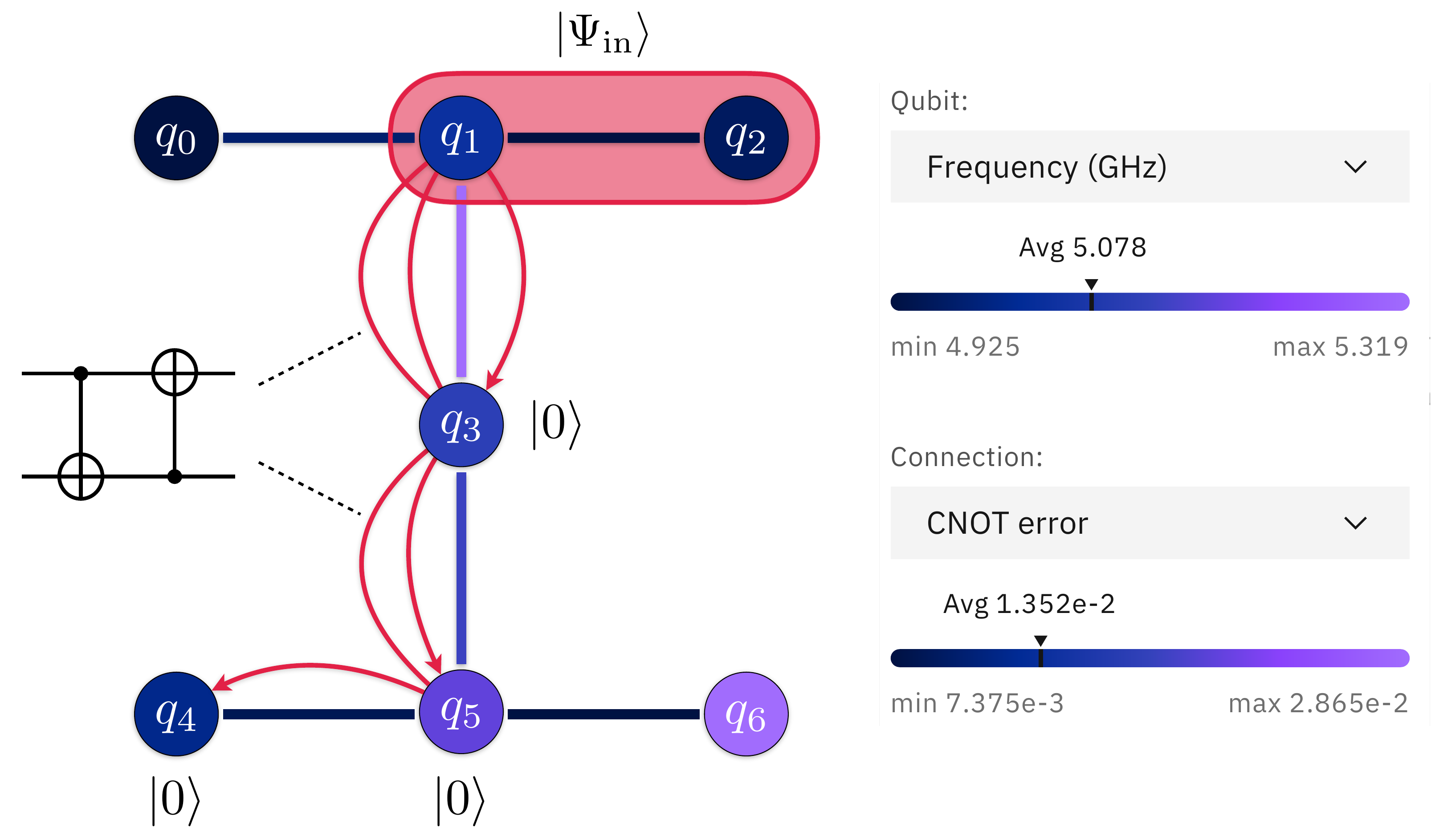}
		\end{centering}
		\caption{Scheme of the \texttt{ibm\char`_oslo} quantum processor experiment (inset is from \url{quantum-computing.ibm.com}).}
		\label{fig:oslo-exp}
\end{figure}

\section{Error suppression in a state transfer}\label{sec:state-transfer}

Here, we consider a practical problem of distributing a two-qubit state over remote physical qubits of a quantum processor.
Specifically, we analyze distributing the entangled state $\ket{\Psi_{\rm in}(\theta=2\pi/3)}$ inside the cloud accessible 7-qubit \texttt{ibm\char`_oslo} quantum processor (see Fig.~\ref{fig:oslo-exp}). 
This can be considered as a prototype experiment for realizing quantum state transfer between distinct quantum information processing devices connected via quantum interface~\cite{Wehner2018}. 

\begin{figure*}[ht!]
    \begin{centering}
		\includegraphics[width=1.33\columnwidth]{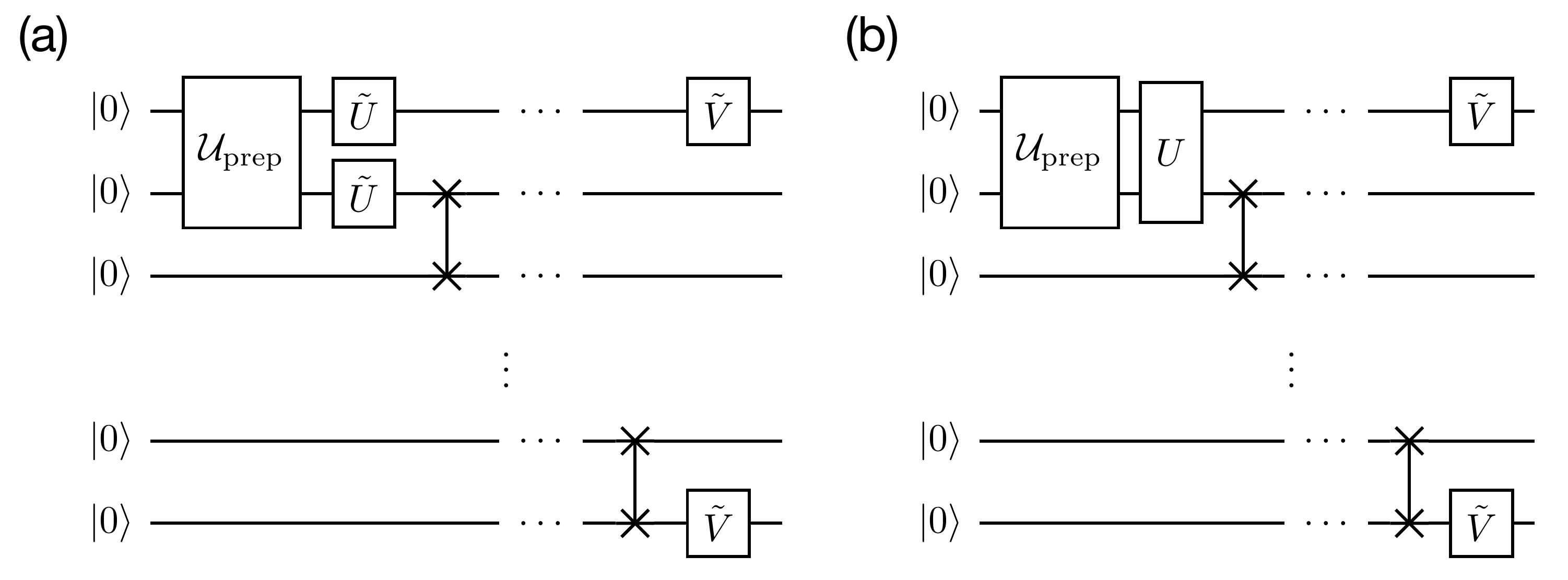}
    \end{centering}
	\caption{Quantum circuits of state transfer experiments with (a) individual-individual protection scheme (b) collective-individual protection scheme.}
	\label{fig:transfer-circs}
\end{figure*}

In this setting, we consider two protection schemes: (i) individual-individual and (ii) collective-individual. 
Quantum circuits of the state transfer experiments are shown in Fig. \ref{fig:transfer-circs}. 
As individual-individual scheme protecting unitaries we take
\begin{equation}
    U = \tilde{U}^{\otimes 2}, \quad \tilde{U} = XH, \quad V=\tilde{V}^{\otimes 2}, \quad \tilde{V} = HX.
\end{equation}
In the case of collective-individual protection scheme, we take
\begin{equation}
    U = (X)^{\otimes 2}{\cal B}^{\dagger}(\xi_{\rm opt}), \quad V=\tilde{V}^{\otimes 2}.
\end{equation}

In our demonstrations, after preparing the input two-qubit state using the $\cal{U}_{\rm prep}$ operator, we transfer the state of one qubit, by $\rm{SWAP}$ operations, through the environment initialized in $\ket{0}$ state. 
Then we reconstruct the whole two-qubit state $\rho_{\rm out}$ by standard linear inversion technique, based on the symmetric informationally complete positive operator-valued measure (SIC-POVM) measurements, on each of two qubits. 
The fidelity is calculated as $\bra{\Psi_{\rm in}}\rho_{\rm out}\ket{\Psi_{\rm in}}$ . 

\begin{figure}[t]
\def\myvdots{\ \vdots\ }
\begin{centering}
\begin{quantikz}[row sep={0.6cm,between origins}, column sep=0.3cm]
\lstick{$\ket{\psi}$}&\qw                    &\qw                   &\targ{}   &\qw      &\meter{} \\ 
\lstick{$\ket{0}$}&\gate[]{R_{y}(\alpha)} & \gate[]{R_{z}(\beta)} &\ctrl{-1} &\gate{H} &\meter{} \\ 
\end{quantikz}
\end{centering}
\vskip -3mm
\caption{Quantum circuit for input state $\ket{\psi}$ tomography.}
\label{fig:tomography-circ}
\end{figure}
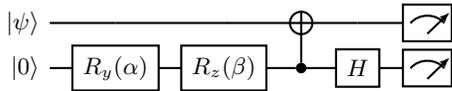

The SIC-POVM quantum state tomography circuit is shown in Fig. \ref{fig:tomography-circ}. 
Tomography circuit utilizes an ancilla qubit for each target qubit.
Corresponding circuit parameters realizing SIC-POVM measurements are given by
\begin{equation}
    \alpha = 2\arccos\left( \sqrt{\frac{1}{2} + \frac{1}{2\sqrt{3}}} \right), \quad \beta = \frac{\pi}{4}.
\end{equation}

Physically, for tomography we use the same environment qubits as during state transfer since after {\rm SWAP} operations, the state of the mediated environment does not change. 
As we provide the real-world example, we optimize the transfer circuit to achieve maximally accessible fidelity.
Therefore, we reduce the {\rm SWAP} operations to two {\rm CNOT} operations thanks to the knowledge of the initial state of the environment.
The topology of the state transfer and optimized operations are shown in Fig. \ref{fig:oslo-exp}. 

\begin{figure*}[ht]
		\begin{centering}
			\includegraphics[width=1.5\columnwidth]{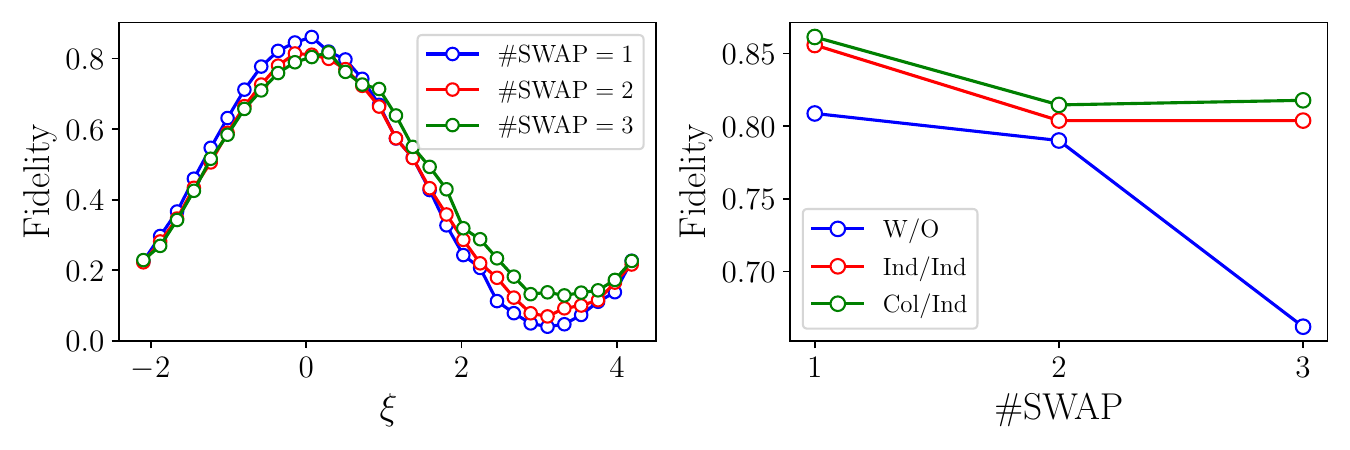}
		\end{centering}
		\caption{Experimental results of state transfer from \textrm{IBM} quantum processor.}
		\label{fig:oslo-res}
\end{figure*}

The result of the optimization of $\xi$ parameter and resulting input-output fidelities are shown in Fig. \ref{fig:oslo-res}. 
We note that as this is a practical example, the experimental data was obtained from transpiled circuits.
We provide the results of several protocol runs corresponding to a different number of {\rm SWAP} operations. 
One can observe a notable increase in fidelity compared to the unprotected case.

    It is worth to mention that the considered approach for a state propagation protection can be used in distributed computing architectures, where two or more `stationary' quantum processors are physically separated and connected by means of `flying' qubits.
    The appearing decoherence channel then consists of imperfections related to interactions between stationary and flying qubits, as well a decoherence during transmission of flying qubits through some medium, e.g. optical fibres or free space.
    Anyway, an individual-individual and a collective-individual schemes can be used in this case, where collective operations are realized within a single quantum processor.

\section{Conclusion}\label{sec:outlook}

We have demonstrated the implementation of the scheme for suppressing decoherence in multi-qubit quantum systems according to the recent theoretical proposal~\cite{Kiktenko2020}.
The simulation of the scheme for suppressing decoherence in two- and four-qubit quantum systems was carried out for three main types of decoherence: 
particularly, depolarizing, dephasing, and amplitude damping, which were presented as quantum channels acting on each qubit of two and four-qubit systems.
As a result, the dependences of the fidelity of the output quantum state on the strength of decoherence have been obtained.
It has been shown that the most advantageous scheme of suppression, as expected, is the collective-collective scheme, which effectively suppresses decoherence in all cases.
Unfortunately, realization of collective operators in the general case requires an exponential number of elementary single- and two-qubit gates, which limits their applicability in realistic conditions.

We have shown the paradigmatic example of decoherence effects suppression in experiments with a cloud-accessible 5-qubit quantum processor \texttt{ibmq\char`_manila}, where the strength of the decoherence is controlled by the delay.
We have observed an increase in the fidelity value both in the case of two-qubit decoherence and in the case of four-qubit decoherence for all types of schemes.
We also have demonstrated the real-world example of quantum state transfer with a cloud-accessible 7-qubit quantum processor \texttt{ibm\char`_oslo}. 
We have observed the increase in fidelity for optimized state transfer protocol up to $10\%$. 
As we expect our findings to be useful for increasing the performance of current NISQ devices.

\section*{Acknowledgements}

We acknowledge use of the IBM Q Experience for this work (our results were collected in September 2022). 
The views expressed are those of the authors and do not reflect the official policy or position of IBM or the IBM Q Experience team.
We thank S. Straupe for useful comments. 
This work was supported by the Russian Roadmap on Quantum Computing (the development of the method and experimental tests; Contract No. 868-1.3-15/15-2021, October 5, 2021),
by the Priority 2030 program at the National University of Science and Technology ``MISIS” under the project K1-2022-027 (analysis of the method), 
and by the Russian Science Foundation Grant No. 19-71-10091 (work of E.K. on Sec.~\ref{sec:state-transfer}, applying the error suppression scheme for quantum state transfer).
The work by ASM is also supported by Scholarship of Russian Federation President  (No. SP-1351.2022.5).

\bibliographystyle{apsrev4-1}
\bibliography{bibliography-rev}
		
\end{document}